\documentclass[epj,nopacs]{svjour}
\usepackage{graphicx}
\usepackage{amsmath}
\newcommand{\half}{{\textstyle\frac{1}{2}}}

\newcommand{\tvec}[1]{\boldsymbol{#1}}

\newcommand{\im}{\operatorname{Im}}
\newcommand{\re}{\operatorname{Re}}

\newcommand{\ms}{\mskip 1.5mu}

\begin{document}
\title{Introduction to GPDs and TMDs}
\author{Markus Diehl
}                     
\institute{Deutsches Elektronen-Synchroton DESY, Notkestr.~85, 22607
  Hamburg, Germany}
\date{ }
%
\abstract{Generalised parton distributions (GPDs) and transverse
  momentum dependent parton distributions (TMDs) describe
  complementary aspects of the three-dimensional structure of hadrons.
  We discuss their relation to each other and recall important theory
  results concerning their properties and their connection with physical
  observables.
} 
\maketitle
\section{Aspects of nucleon structure}
\label{sec:intro}

Among the most intriguing aspects of quantum chromodynamics is the stark
contrast between the simplicity of its Lagrangian, formulated in terms of
quark and gluon fields, and the complexity of its bound states, hadrons.
To understand hadron structure -- in particular the structure of the
nucleon -- in terms of quarks and gluons remains among the most challenging
tasks in particle physics.  It is also of acute practical importance for
the quantitative interpretation of high energy experiments, first and
foremost at the Large Hadron Collider.

The parton model describes a fast moving nucleon as a collection of
quasi-free quarks, antiquarks and gluons, whose longitudinal momentum
distribution is described by parton densities.  The formalism of collinear
factorisation implements these ideas in QCD and provides the backbone of
phenomenology in hadron-hadron and lepton-hadron collisions.  In several
situations it does however not adequately capture all relevant physics,
and the conventional parton densities quantify nucleon structure only in a
single space dimension.

The distribution of a parton in the plane transverse to the direction of
motion of its parent nucleon involves two complementary aspects:
\begin{enumerate}
\item the transverse momentum of a parton in the nucleon, quantified by
  TMDs (transverse momentum dependent distributions), leaves its imprint
  on the transverse momenta of particles in the final state and can thus
  be accessed by suitable experimental observables.
\item the transverse position of a parton in the nucleon can be accessed
  in suitable exclusive processes where the proton is deflected at small
  angles.  The transverse spatial distribution of the parton can be
  reconstructed by a Fourier transform from the transverse momentum
  transferred to the proton, in close analogy with the principle of
  $X$-ray crystallography.  Generalised parton distribution (GPDs)
  quantify this type of information.
\end{enumerate}
In both cases, the relevant length or momentum scales are typical of
non-perturbative dynamics, in contrast to the dimensionless longitudinal
momentum \emph{fractions} on which all types of parton distributions
depend.  In this sense, TMDs and GPDs carry in a more direct way the
imprint of non-perturbative phenomena such as confinement.

Due to the uncertainty principle, transverse momentum and position (along
the same direction) cannot be specified simultaneously.  However, a
unifying description of both aspects can be given using Wigner
distributions.  These different degrees of freedom and their interplay are
discussed in section~\ref{sec:variables}.

The discussion of ``nucleon structure'' at the level of quarks and gluons
is closely related to the dynamics of the physical processes in which this
structure is ``probed''.  To connect nucleon structure with experimental
observables in a controlled way, we rely on factorisation formulae --
proven or conjectured.  These formulae express an observable in terms of
parton distributions or similar quantities involving non-perturbative
dynamics on the one hand and parton-level cross sections or similar
quantities involving only dynamics at short distances on the other hand.
The latter can be computed in QCD perturbation theory.

In particular the discussion of TMDs shows that the separation of
``structure'' and ``probe'' can be surprisingly intricate, to the point
of challenging a naive picture of ``nucleon structure'' itself.
Theoretical work aimed at showing exactly when and how information on
nucleon structure can be extracted from experimental observables has
significantly brought (and continues to push) forward our understanding of
QCD in several areas, such as the scale evolution of operator matrix
elements, the control of power suppressed (higher twist) contributions,
the physical relevance of ``unphysical'' gluon polarisation, the
transition between perturbative and non-perturbative regimes, and the
limits of applicability of factorisation.  We give a brief overview of
such aspects for GPDs in section~\ref{sec:GPDs} and for TMDs in
section~\ref{sec:TMDs}.

As GPDs and TMDs contain information about the longitudinal and at least
one of the two transverse directions, they have a much richer spin
dependence than the conventional parton densities (the spin dependence of
which is strongly restricted by rotational invariance).  Several GPDs and
TMDs describe specific spin-orbit correlations at the parton level and are
sensitive to parton orbital angular momentum, which is a crucial
ingredient in describing how the overall spin of the nucleon arises from
its constituents.  In section~\ref{sec:spin} we make some comments on this
topic, which is reviewed in detail in a dedicated contribution to this
volume~\cite{Liu:2015xha}.

For definiteness, we will mostly consider distributions for quarks and
antiquarks in the following.  Gluon distributions can be discussed in
close analogy, with appropriate adaptions.


\section{Space-time and momentum structure}
\label{sec:variables}

In this section we review the variables on which different kinds of parton
distributions depend.  This will allow us to see how the different
distributions are related to each other.  Any process that probes partons
inside a nucleon singles out a particular direction, providing a physical
distinction between ``longitudinal'' and ``transverse''.  This is
naturally implemented in the parton model, where one chooses a reference
frame in which the hadron under consideration moves fast.  One is however
not limited to this choice: parton distributions are defined in a
covariant way, and one can also discuss them in the hadron rest frame.  Of
course, the process probing the parton still singles out a particular
direction in that frame, so that transverse and longitudinal directions
play different roles.  Thus, the information one can gain about partons
in the proton inevitably breaks manifest three-dimensional rotation
invariance.  For definiteness, we will in the following consider a
reference frame in which the hadron moves fast in the positive $z$
direction (exactly or approximately).  A suitable set of coordinates is
then given by the light-cone coordinates $v^\pm = (v^0 \pm v^3) /\sqrt{2}$
and the transverse components $\tvec{v} = (v^1, v^2)$ of a given
four-vector $v$.

\begin{figure}[hb]
\begin{center}
  \includegraphics[width=0.4\textwidth]{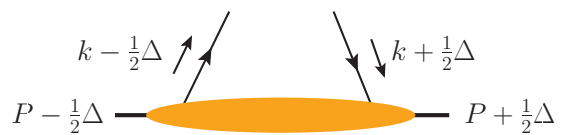}
\end{center}
\caption{\label{fig:correlator} Momentum assignments in the general
  quark correlation function \protect\eqref{corr}.}
\end{figure}

A two-parton correlation function for quarks is defined as the matrix
element of a bilinear quark field operator between proton states:
\begin{align}
  \label{corr}
  & H(k, P, \Delta) = (2\pi)^{-4}
    \int d^4z\; e^{izk}
    \nonumber \\
  &\quad \times \bigl\langle p(P + \half\Delta) |
    \bar{q}(-\half z) \ms\Gamma\ms q(\half z) | p(P - \half\Delta)
    \bigr\rangle \,.
\end{align}
The Dirac matrix $\Gamma$ selects the twist\footnote{There are several --
  slightly different -- definitions of the term ``twist''.  We will not
  expand on this topic here and refer to \protect\cite{Jaffe:1996zw} for a
  detailed discussion.}
and the parton spin degrees of freedom, and we have omitted labels for the
proton spin state.  For the moment we put aside field theoretical issues
such as the regularisation and renormalisation of the operator and the
insertion of a Wilson line between the two quark quark fields.  The parton
and proton momenta are shown in figure~\ref{fig:correlator}.  Notice that
the on-shell condition for the proton states results in the conditions $P
\Delta = 0$ and $4 P^2 + \Delta^2 = 4 m^2$, where here and in the
following $m$ denotes the proton mass.

While $H(k, P, \Delta)$ is a smooth function of $\Delta$, the cases where
this momentum transfer is zero or not correspond to distinct physical
situations:
\begin{enumerate}
\item In the forward limit $\Delta=0$ the function appears in the
  \emph{cross section} of inclusive processes.  Glossing over
  complications from confinement, one may insert a complete set $|X\rangle
  \langle X|$ of states between the fields $\bar{q}$ and $q$ in the matrix
  element \eqref{corr}.  This gives essentially the amplitude
  $\mathcal{A}$ for emitting a quark or antiquark from the proton, with a
  system of spectator partons $X$ left behind, multiplied by the conjugate
  $\mathcal{A}^*$ of that amplitude as required for the computation of a
  cross section.  The representation as a squared amplitude $\mathcal{A}^*
  \mathcal{A}$ opens the possibility to interpret certain forward
  distributions as probability densities in the sense of quantum
  mechanics.  Taken literally, this interpretation no longer holds after
  the regularisation and renormalisation already mentioned, but if taken
  with due caution it remains a valuable guide for physical intuition.

  We note that in the forward limit, it is convenient to take a frame
  where $\tvec{P}=\tvec{0}$, so that the proton moves exactly along the
  $z$ axis.
\item In non-forward kinematics $\Delta\neq 0$ the function appears in
  the \emph{amplitude} of exclusive reactions, with an incoming proton
  of momentum $P-\Delta/2$ and an outgoing one of momentum $P+\Delta/2$.
  The functions in this case are often called ``generalised''.
\end{enumerate}
In physical observables, the correlation function \eqref{corr} typically
is integrated over one or more components of the four-momentum $k$.  Let
us review this step by step.
\begin{enumerate}
\item After an integral over $k^-$, the quark and antiquark fields are
  evaluated at $z^+ = 0$.  This admits a very elegant interpretation in
  the framework of light-cone quantisation: quark fields are quantised at
  light-cone time $z^+ = 0$, where they obey the anticommutation relations
  for free fields and have a Fourier decomposition in terms of creation
  and annihilation operators for quarks and antiquarks.  This may be seen
  as the field theory implementation of the parton model, where partons
  are regarded as quasi-free just before they are probed in a physical
  process.  The parton states created or annihilated by the fields have
  positive plus-momentum, so that depending on the respective signs of
  $k^+ - \Delta^+ /2$ and $k^+ + \Delta^+ /2$, the matrix element in
  figure~\ref{fig:correlator} describes the emission and reabsorption of a
  quark, the emission and reabsorption of an antiquark, or (for $\Delta^+
  \neq 0$ only) the emission or absorption of a quark-antiquark pair (see
  figure~\ref{fig:gpd-bspace} below).  At $z^+=0$, the representation of
  the parton correlation function as product $\mathcal{A}^* \mathcal{A}$
  of an amplitude and its conjugate turns into a product $\psi^* \psi$ of
  a light-cone wave function and its conjugate.  More detail is given in
  the contribution \cite{Burkardt:2015qoa} to this volume.

  Note that after integration over $k^-$ the partons no longer have a
  definite virtuality, in particular they are \emph{not} on their mass
  shell.  When computing the hard parton-level subprocess for a
  factorisation formula, one may indeed neglect the parton virtuality
  (which is much smaller than the hard scale of the process, given
  e.g.\ by the photon virtuality in deep inelastic scattering or Drell-Yan
  production).  From the point of view of proton structure, however, the
  virtuality of partons is essential.  In fact, the notion of on-shell
  partons is at odds with the phenomenon of confinement.

  In a field theoretic setting, the integration over $k^-$ can induce so
  called rapidity divergences in the matrix element \eqref{corr}, which
  come from regions where gluons inside a right-moving proton have
  infinite negative rapidity.  Such regions are naturally cut off in a
  physical process, but in the matrix element they must be excluded by a
  suitable regulation procedure.  The distributions are then dependent on
  a parameter acting like a rapidity cutoff, and the dependence on this
  parameter is described by the Collins-Soper equation.  We will come back
  to this in section~\ref{sec:TMDs}.
\item The distributions obtained after an additional integral over
  $\tvec{k}$ are often called ``collinear''.  This does of course not
  imply that the partons in the proton have no transverse momentum, but
  rather that one is not sensitive to their transverse momentum in
  observables that involve collinear distributions.

  The combined integral $\int dk^-\, d^2\tvec{k}$ puts the separation
  between the fields in \eqref{corr} on the light-cone, $z^2=0$, which
  leads to important simplifications in the field theoretic context.  The
  rapidity divergences mentioned in the previous point disappear after
  integrating over $\tvec{k}$, so that a rapidity cutoff parameter is no
  longer needed.  In turn, the $\tvec{k}$ integration leads to ultraviolet
  divergences in the matrix element, which need to be renormalised.  They
  come from regions where partons have infinitely large transverse
  momentum and virtuality.  The dependence of the collinear distribution
  on the ultraviolet renormalisation scale $\mu$ is described by the
  familiar DGLAP evolution equations, or by generalised versions of these
  in the case $\Delta^+ \neq 0$.
\item If one also integrates over $k^+$, then the parton momentum is fully
  unspecified and the product of fields in the matrix element \eqref{corr}
  becomes a local current.  The matrix element is then described by one or
  more form factors, which depend only on the invariant momentum transfer
  $\Delta^2$ due to Lorentz invariance.

  If the current is conserved, the dependence on the renormalisation
  scale disappears; otherwise it is described by a simple
  renormalisation group equation with an anomalous dimension depending
  on the current.
\end{enumerate}
So far our discussion was purely in momentum space.  Let us now see how
different momentum components in \eqref{corr} can be traded for position
space arguments by a Fourier transform.  Again, the transverse and
longitudinal directions play very different roles, and for now we focus on
the transverse ones.  Starting with momentum eigenstates $|p(P^+,
\tvec{P})\rangle$ of the proton, one can form wave packets
\begin{align}
  \label{imp-state}
|p(P^+, \tvec{b})\rangle & \propto \textstyle\int d^2\tvec{P}\ e^{-i
    \tvec{P}\tvec{b}}\, |p(P^+,\tvec{P})\rangle \,,
\end{align}
which are localised at position $\tvec{b}$ in the transverse plane.  They
are indeed eigenstates of a transverse position operator
\cite{Soper:1972xc}, which means that a relativistic particle of mass $m$
can be localised \emph{exactly} in two dimensions.  (In all three
dimensions, this is only possible up to ambiguities of the order of its
Compton wave length $1/m$).

To understand what this localisation means for a spatially extended
object like the proton, we consider so-called transverse boosts, which
form a subgroup of the Lorentz group and transform a vector $k$ as
\begin{align}
  \label{transv-boosts}
k^+ & \to k^+ \,, & \tvec{k} & \to \tvec{k} - k^+ \tvec{v} \,,
\end{align}
where $\tvec{v}$ is a fixed vector characterising the transformation.
Notice the analogy between these transverse boosts and the familiar
Galilean transformations in nonrelativistic mechanics, which are obtained
from \eqref{transv-boosts} by replacing $k^+$ with the mass $m$ of the
particle.  The relativistic analogue of the centre-of-mass for a composite
system is thus the centre of plus-momentum, $\tvec{b} = \sum_i k_i^+
\tvec{b}_i^{} /\sum_i k_i^+$.  This is the meaning of $\tvec{b}$ in
\eqref{imp-state}.

As already mentioned, a quark field operator $q(z^-, \tvec{z})$ at $z^+=0$
can be decomposed on annihilation and creation operators for parton
states.  From matrix elements of bilinear field operators at $z^+=0$
between the proton states in \eqref{imp-state} we can thus define
so-called impact parameter distributions, which describe the transverse
spatial distribution of partons inside a proton that is localised in the
transverse plane \cite{Burkardt:2000za}.  These distributions are obtained
from their momentum space counterparts by a two-dimensional Fourier
transformation w.r.t.\ the momentum transfer $\tvec{\Delta}$.

Let us take a closer look at the relation between transverse momentum and
position variables.  The Fourier transform
\begin{align}
q(z^-, \tvec{k}) &= \textstyle\int d^2\tvec{z}\, e^{- i
    \tvec{k}\tvec{z}}\, q(z^-, \tvec{z}) \big|_{z^+=0} \,,
\end{align}
contains the annihilation operator for a quark with transverse
momentum $\tvec{k}$, and for the bilinear operator needed to form a
quark density we have
\begin{align}
& \bar{q}(z_2^-, \tvec{k}_2^{}) \ms\Gamma\ms q(z_1^-, \tvec{k}_1^{})
\nonumber \\
 &\quad = \textstyle\int d^2\tvec{z}_2^{}\, d^2\tvec{z}_1^{}\,
    e^{i (\tvec{k}_2 \tvec{z}_2 - \tvec{k}_1 \tvec{z}_1)}\,
    \bar{q}(z_2^-, \tvec{z}_2^{}) \ms\Gamma\ms q(z_1^-, \tvec{z}_1^{}) \,.
\end{align}
Rewriting the Fourier exponent as
\begin{align}
\tvec{k}_2 \tvec{z}_2 - \tvec{k}_1 \tvec{z}_1
 &= \half (\tvec{z}_2 + \tvec{z}_1) (\tvec{k}_2 - \tvec{k}_1)
\nonumber \\
 &\quad + \half (\tvec{k}_2 + \tvec{k}_1) (\tvec{z}_2 - \tvec{z}_1) \,,
\end{align}
we can read off the relation between Fourier conjugate variables:
\begin{align*}
\text{average position} & \leftrightarrow \text{momentum difference} \,,
\\
\text{average momentum} & \leftrightarrow \text{position difference} \,,
\end{align*}
where ``average'' and ``difference'' refer to the right and left hand
sides of figure~\ref{fig:correlator}, or equivalently to the light-cone
wave function $\psi$ and its conjugate $\psi^*$.

\begin{figure*}
  \includegraphics[width=0.85\textwidth]{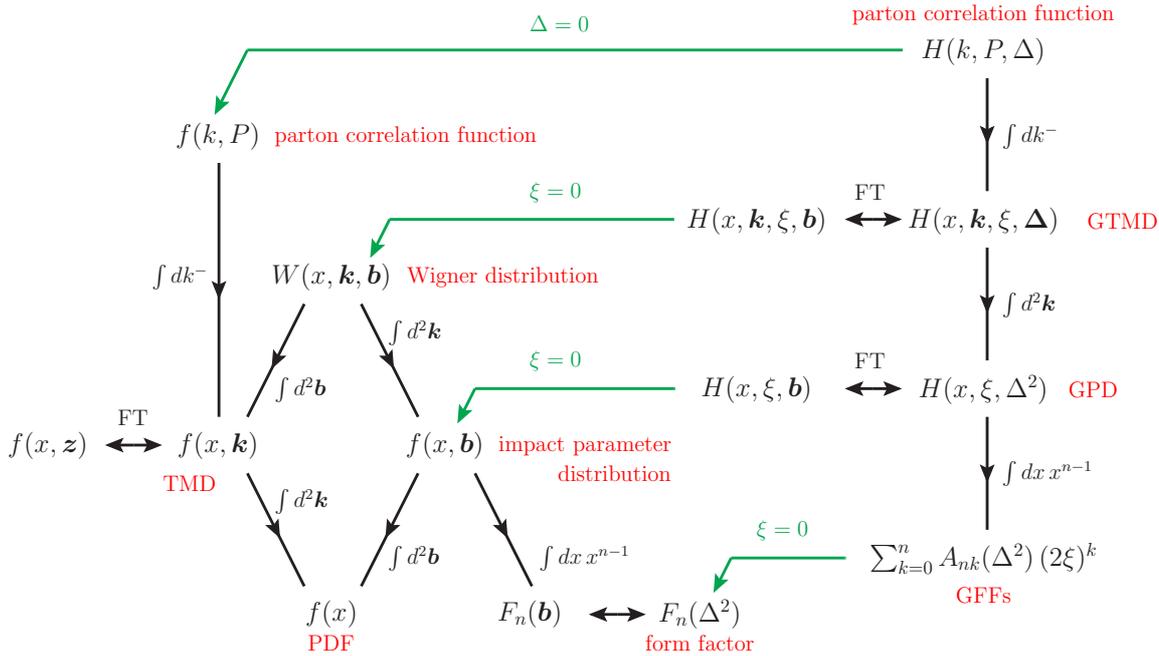}
  \caption{\label{fig:scheme} Selected quantities that can be derived
    from the fully differential two-quark correlation function
    $H(k,P,\Delta)$ defined in \protect\eqref{corr}.  Double arrows
    marked by ``FT'' denote a Fourier transform between
    $\tvec{\Delta}$ and $\tvec{b}$ or between $\tvec{k}$ and
    $\tvec{z}$.  Fractions of plus-momentum (commonly called
    ``longitudinal momentum fractions'') are written as $x = k^+/P^+$
    and $2\xi = -\Delta^+/P^+$.  The invariant momentum transfer can
    be expressed in terms of longitudinal and transverse variables as
    $\Delta^2 = - (4\xi^2 m^2 + \tvec{\Delta}^2) /(1-\xi^2)$.
    Only kinematic arguments of the functions are given, while the
    scales introduced by ultraviolet renormalisation ($\mu)$ of by the
    regulation of rapidity divergences ($\zeta$) are suppressed.  As
    discussed in the text, the integrals $\int dk^-$ and
    $\int d^2\tvec{k}$ cannot be taken literally but must be
    supplemented with a regularisation procedure.}
\end{figure*}

After these general considerations, we can take a closer look at the
different distributions that can be obtained from the general
two-quark correlation function in \eqref{corr}.  A selection of them
is shown in figure~\ref{fig:scheme}.  Let us start at the top of the
hierarchy.
\begin{enumerate}
\item In the forward limit $\Delta=0$, parton correlation functions that
  are not integrated over any component of $k$ (called ``doubly'' or
  ``fully unintegrated'' distributions) have been discussed in the context
  of evolution at small $x$ \cite{Watt:2003mx} and with the aim of having
  an exact description of final-state kinematics
  \cite{Collins:2005uv,Collins:2007ph}.  Under the name of ``beam
  functions'', they have also been introduced in soft-collinear effective
  theory (SCET) for the resummation of large logarithms in observables
  sensitive to the proton remnants (called ``beam jets'')
  \cite{Stewart:2009yx,Stewart:2010qs,Jain:2011iu}.  In that case,
  distributions differential in $k^-$ but integrated over $\tvec{k}$ are
  referred to as beam functions as well.  The considerations in
  \cite{Watt:2003mx} and \cite{Stewart:2009yx,Stewart:2010qs,Jain:2011iu}
  focus on the region of large parton virtuality $k^2$ and compute the
  unintegrated distributions in terms of conventional parton distribution
  functions (PDFs), an aspect we will discuss in more detail for TMDs in
  section~\ref{sec:TMDs}.

  A detailed analysis of factorisation with unintegrated distributions has
  been given for semi-inclusive deep inelastic scattering (SIDIS) in
  \cite{Collins:2007ph}.  For hadron-hadron collisions there are strong
  arguments that this type of factorisation generically fails, due to soft
  gluon exchange between the spectator partons in each hadron
  \cite{Gaunt:2014ska,Zeng:2015iba}.  In kinematics referred to as the
  Glauber region, these soft interactions ``tie together'' the two
  had\-rons in a way that prevents one from describing the
  non-perturbative dynamics by matrix elements that pertain to only one
  hadron and not to both.  To establish factorisation, one has to show
  that (after appropriate approximations) gluon exchange in the Glauber
  region cancels in the observable at hand.

  Not being integrated over any momentum component, parton correlation
  functions retain manifest Lorentz invariance (provided that one is
  careful not to forget auxiliary vectors required for their field
  theoretical definition).  They can therefore be used to classify and
  relate different distributions that descend from them.  Examples are
  given in \cite{Musch:2010ka} for $\Delta=0$ and in
  \cite{Meissner:2009ww} for $\Delta\neq 0$.
\item Wigner distributions depend on the average momentum and the
  average position of the quark.  From the uncertainty principle it is
  clear that they cannot represent joint probabilities in these two
  variables, but integrating over any one of them, one obtains a
  probability in the other.

  The most straightforward interpretation of these distributions is in
  the forward limit $\xi=0$ of longitudinal momentum.  Integrating the
  Wigner distribution $W(x,\tvec{k},\tvec{b})$ in
  figure~\ref{fig:scheme} over $\tvec{b}$ one obtains the TMD
  $f(x,\tvec{k})$, which specifies the probability density of finding
  a parton with longitudinal momentum fraction $x$ and transverse
  momentum $\tvec{k}$, whilst integration over $\tvec{k}$ yields the
  impact parameter distribution $f(x,\tvec{b})$, which gives the
  probability density of finding a parton with longitudinal momentum
  fraction $x$ at a transverse distance $\tvec{b}$ from the centre of
  momentum of the proton.

  There is no process known where the Wigner distributions just
  described would be directly accessible to experiment.  It is amusing
  to note that more complicated Wigner distributions do appear in the
  cross section for double parton scattering -- a process where in a
  single proton-proton collision two pairs of partons initiate two
  separate hard scatters \cite{Diehl:2011tt}.  These double parton
  distributions
  $F(x_1,x_2,\tvec{k}_1,\tvec{k}_2,\tvec{b}_1,\tvec{b}_2)$ are defined
  in analogy to $W(x,\tvec{k},\tvec{b})$, but with two bilinear quark
  operators instead of one.  The cross section involves the product
  \begin{align}
    F(x_1,x_2,\tvec{k}_1,\tvec{k}_2,\tvec{b}_1,\tvec{b}_2)\,
    F(\bar{x}_1,\bar{x}_2,\bar{\tvec{k}}_1,\bar{\tvec{k}}_2,
      \bar{\tvec{b}}_1,\bar{\tvec{b}}_2)
  \end{align}
  of two such distributions, one for each proton.  It is integrated
  over the transverse variables with constraints
  $\tvec{k}_1 + \bar{\tvec{k}}_1 = \tvec{q}{}_1$ and
  $\tvec{k}_2 + \bar{\tvec{k}}_2 = \tvec{q}{}_2$, where $q_1$ and
  $q_2$ are the momenta of the particles produced in one and the other
  hard scatter.  The additional constraint
  $\tvec{b}_1 - \tvec{b}_2 = \bar{\tvec{b}}_1 - \bar{\tvec{b}}_2$
  ensures that the two parton pairs in each proton can initiate two
  separate short-distance processes.

  Because Wigner distributions do not represent probabilities, they need
  not be positive, which makes an intuitive interpretation somewhat
  difficult.  Using an appropriate smearing procedure, one can obtain
  so-called Husimi distributions, which are positive definite and can be
  interpreted as probabilities without contradicting the uncertainty
  principle.  They have been discussed in the context of nucleon structure
  in \cite{Hagiwara:2014iya}.
\item The preceding discussion referred to Wigner distributions that
  depend on momenta and positions in the transverse plane whilst in
  the longitudinal direction the momentum representation is kept.
  This ``mixed'' representation is useful in many contexts (for
  instance in small-$x$ physics), and it allows one to keep a close
  connection to the parton model picture, where hadrons and partons
  move fast in the longitudinal direction.  However, one may also
  Fourier transform longitudinal momentum variables.
  Three-dimensional Wigner functions have been defined in
  \cite{Belitsky:2003nz} by a Fourier transform of GTMDs (see
  figure~\ref{fig:scheme}) w.r.t.\ the momentum transfer
  $(\Delta^1, \Delta^2, \Delta^3)$ in the brick-wall frame, where
  $\Delta^0=0$.  The interpretation of these quantities is similar to
  the familiar interpretation of Fourier transformed form factors as
  three-dimensional spatial densities \cite{Sachs:1962zzc}, and it is
  subject to the same limitations due to special relativity (see our
  comment above).  A different three-dimensional representation has
  been proposed in \cite{Brodsky:2006ku}, considering the scattering
  amplitude $\mathcal{A}(\tvec{\Delta},\xi)$ for deeply virtual
  Compton scattering (which is closely related to GPDs) and performing
  a Fourier transform w.r.t.\ $\tvec{\Delta}$ and $\xi$.  The
  resulting spatial distribution in the longitudinal direction is
  reminiscent of an optical diffraction pattern.
  
  The Fourier transform w.r.t.\ the average momentum fraction $x$ has
  been studied in particular for distributions integrated over
  $\tvec{k}$, that is for GPDs and PDFs.  It gives matrix elements of
  operators with all fields separated along the light-cone, whose
  ultraviolet renormalisation can be discussed independently of the
  hadron momenta.  This provides deep insight into the apparently
  different scale evolution of PDFs and of meson distribution
  amplitudes, which are matrix elements of the same operators between
  a meson state and the vacuum.  The evolution equations relevant for
  GPDs turn out to interpolate between these two cases
  \cite{Mueller:1998fv,Belitsky:2005qn}.

\item TMDs can be measured in a variety of reactions in lepton-proton
  and proton-proton collisions, where a final-state particle is
  observed with a transverse momentum much smaller than the hardest
  scale in the process (such as the photon virtuality in SIDIS or
  Drell-Yan production).  The measured transverse momentum typically
  results from the convolution of two transverse-momentum dependent
  quantities, such as a parton distribution and a fragmentation
  function in SIDIS or two parton distributions in the Drell-Yan
  process.

\begin{figure*}[t]
\begin{center}
  \includegraphics[width=0.45\textwidth]{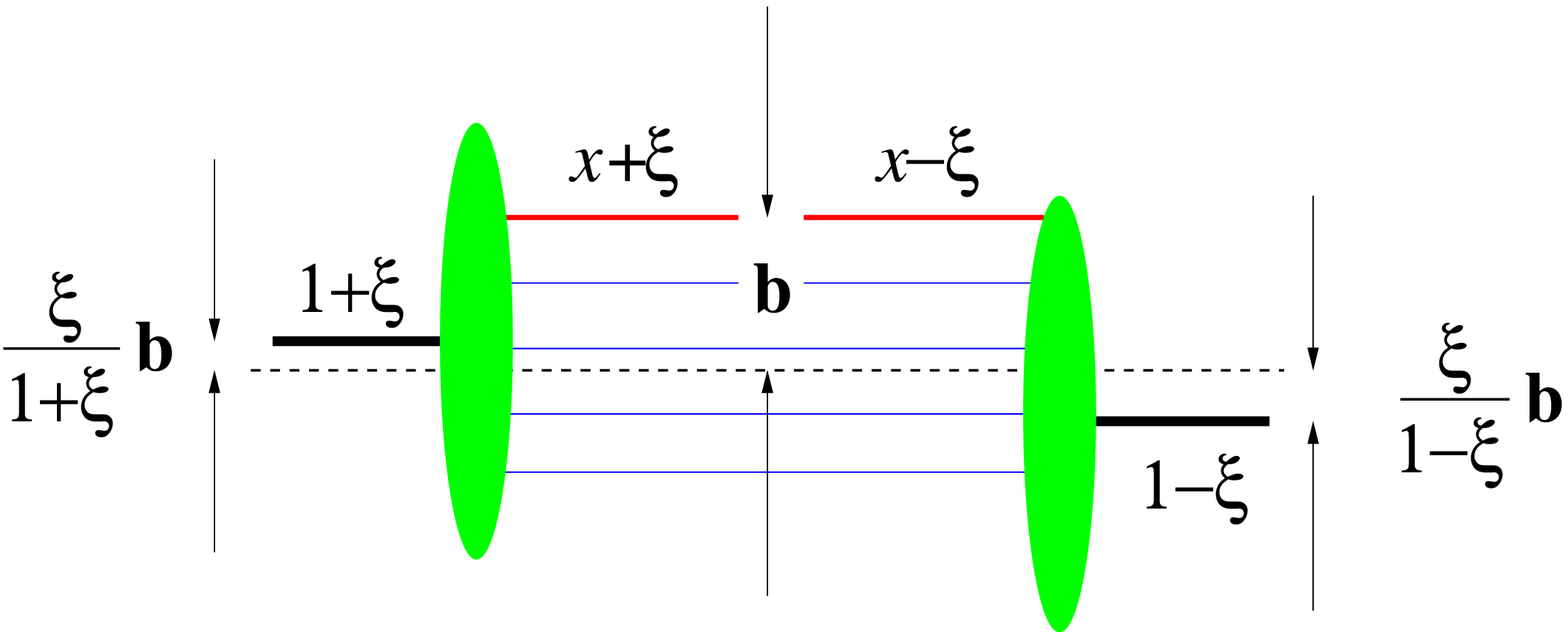}
\hspace{3em}
  \includegraphics[width=0.45\textwidth]{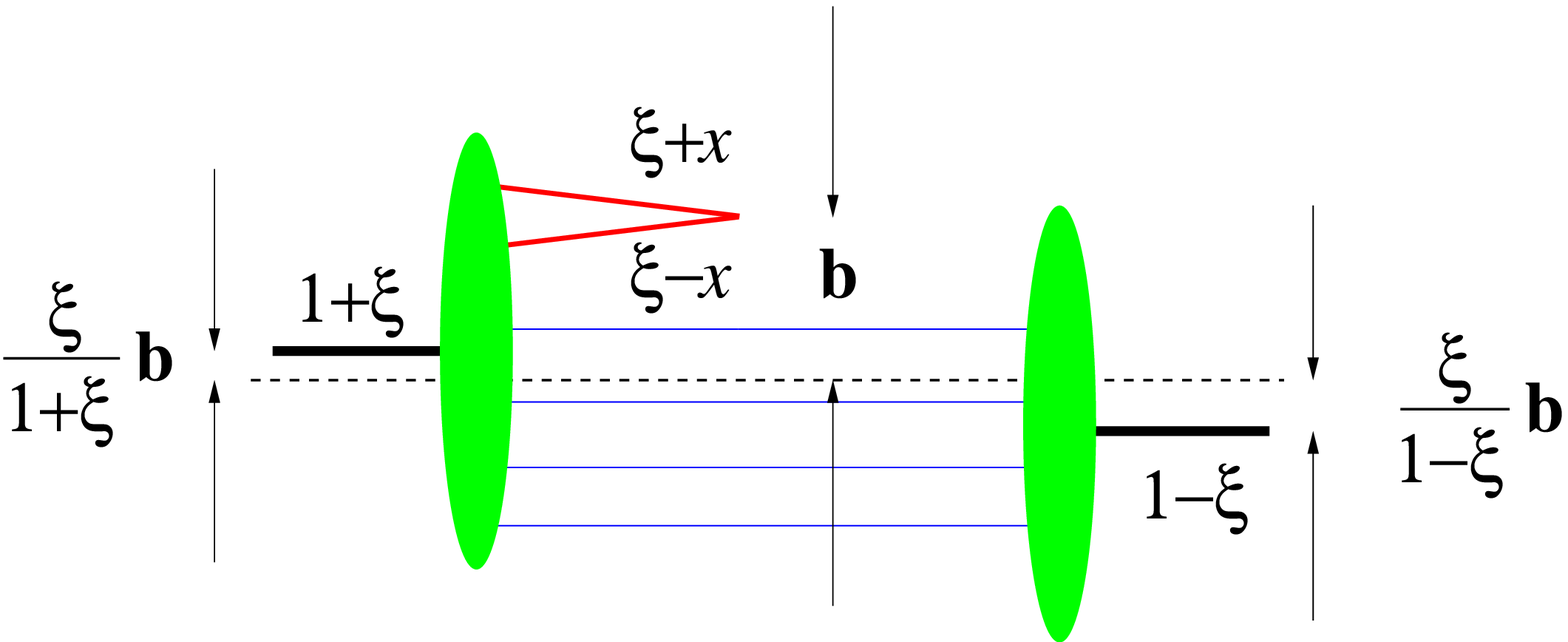}
  \caption{\label{fig:gpd-bspace} Impact parameter space
    representation of a GPD at nonzero skewness variable $\xi$ in the
    regions $\xi<x<1$ (left) and $|x|<\xi$ (right).  $\xi$ is taken to
    be positive, as appropriate for the processes in which GPDs are
    known to occur.  The overall centre of plus-momentum shifts
    because of the transfer of plus momentum to the proton.  The case
    $-1<x<-\xi$ (not shown) is analogous to $\xi<x<1$, with an
    antiquark carrying momentum fraction $-x+\xi$ in the wave function
    and momentum fraction $-x-\xi$ in the conjugate wave function.}
\end{center}
\end{figure*}

  This transverse-momentum convolution becomes an ordinary product
  after a Fourier transform; the resulting distribution
  $f(x,\tvec{z})$ in figure~\ref{fig:scheme} is often called
  $f(x,\tvec{b})$ in the literature (a notation reserved for impact
  parameter distributions here).  This representation has also the
  advantage that Collins-Soper evolution in rapidity (see above) is at
  fixed $\tvec{z}$, whereas in transverse-momentum space it involves
  again a convolution.  Notice that $f(x,\tvec{z})$ does not have a
  simple probability interpretation, $\tvec{z}$ being the difference
  between the parton position in the light-cone wave function and its
  conjugate according to our analysis above.
\item GPDs $H(x,\xi,\Delta^2)$ are accessible in suitable exclusive
  processes, in which the invariant momentum transfer $\Delta^2$ to
  the proton is much smaller than the hard scale $Q^2$, which is often
  given by the virtuality of a photon.  Example processes are deeply
  virtual Compton scattering (DVCS), $\gamma^* p\to \gamma\ms p$, and
  exclusive meson production $\gamma^* p\to M p$, both measurable in
  lepton-proton collisions.  A close analogue of DVCS is timelike
  Compton scattering (TCS), $\gamma\ms p\to \gamma^* p$, where the
  timelike photon decays into a lepton-antilepton pair.  Whereas
  $\Delta$ is directly measurable, the average parton momentum
  fraction $x$ always appears convoluted with a function representing
  the hard-scattering process.  The corresponding deconvolution
  problem is perhaps the most difficult aspect in extracting
  information about GPDs from experimental data.

  The analogous functions that are also differential in the transverse
  momenta of the partons are called generalised TMDs (GTMDs).  Whilst
  the appearance of GPDs in exclusive scattering amplitudes can be
  established with great theoretical rigour
  \cite{Collins:1996fb,Collins:1998be}, GTMDs have only been used in a
  phenomenological spirit, in order to quantify subleading corrections
  in $1/Q$ to DVCS and exclusive meson production
  \cite{Vanderhaeghen:1999xj}.  It is not clear to which extent this
  could be made more rigorous, and even in the setting just mentioned,
  the average transverse parton momentum $\tvec{k}$ appears under an
  integral and cannot be reconstructed in a direct manner.

\item A Fourier transform of GPDs (or GTMDs) w.r.t.\ the transverse
  part $\tvec{\Delta}$ of the momentum transfer yields distributions
  depending on the impact parameter $\tvec{b}$, which quantifies the
  distance of a transversely localised parton from the centre of
  momentum of the proton.  The exclusive processes in which GPDs (or
  GTMDs) appear involve a finite longitudinal momentum transfer to the
  proton, so that the corresponding distributions are probed at
  nonzero $\xi$.  This complicates somewhat the interpretation of
  $H(x,\xi,\tvec{b})$, which has interference character as far as the
  longitudinal momentum fractions are concerned, as shown in
  figure~\ref{fig:gpd-bspace}.  Only when one takes the limit $\xi=0$
  of this distribution does one obtain the impact parameter
  distribution $f(x,\tvec{b})$ with a density interpretation.
  Likewise, one obtains the Wigner distribution
  $W(x,\tvec{k},\tvec{b})$ as the $\xi=0$ limit of a Fourier
  transformed GTMD.

\item Integrating TMDs over $\tvec{k}$ or impact parameter
  distributions over $\tvec{b}$, one obtains the usual collinear PDFs.
  Among all quantities discussed so far, the range of processes where
  these distributions can be studied experimentally is largest.
  Correspondingly, our knowledge about them is very advanced and often
  in the realm of precision physics.  However, even for PDFs some
  aspects remain poorly known, such as the distributions of strange
  quarks and antiquarks or the distribution of longitudinally
  polarised gluons.
\item Taking Mellin moments $\int dx\, x^{n-1}$ of a GPD, one obtains
  a sum over the form factors of a local current, which are often
  called generalised form factors (GFFs) in this context.  As a
  consequence of Lorentz invariance, they are weighted by powers of
  $\xi$ in the sum, which strongly constrains the interplay between
  the $x$ and $\xi$ dependence of the original GPD.  This
  polynomiality property plays a prominent role in the construction of
  consistent GPD parametrisations.

  Hadron matrix elements of local currents in the spacelike region can
  readily be continued to Euclidean spacetime and thus be computed in
  lattice QCD.  This opens the way to obtain genuinely
  non-perturbative information about GPDs from first principles and is
  complementary to model building efforts.  A detailed review of this
  field is given in \cite{Hagler:2009ni}.

  Evaluating the Mellin moments of GPDs in the $\xi=0$ limit and
  performing a two-dimensional Fourier transform to impact parameter
  space, one obtains the Mellin moments of the impact parameter
  distribution $f(x,\tvec{b})$, as seen in figure~\ref{fig:scheme}.
  One can thus interpret elastic form factors as
  \emph{two-dimensional} densities in the impact parameter plane, with
  the information about the longitudinal parton momentum being
  condensed into an average with weight $x^{n-1}$.  This provides a
  parton-based alternative to the representation of form factors as
  Fourier transformed three-dimensional densities (see above) and has
  led to surprising insights when applied to the electromagnetic
  proton and neutron form factors, which are rather well known
  experimentally \cite{Miller:2007uy}.
\end{enumerate}


\section{More on GPDs}
\label{sec:GPDs}

\begin{figure*}
\begin{center}
  \includegraphics[width=0.25\textwidth]{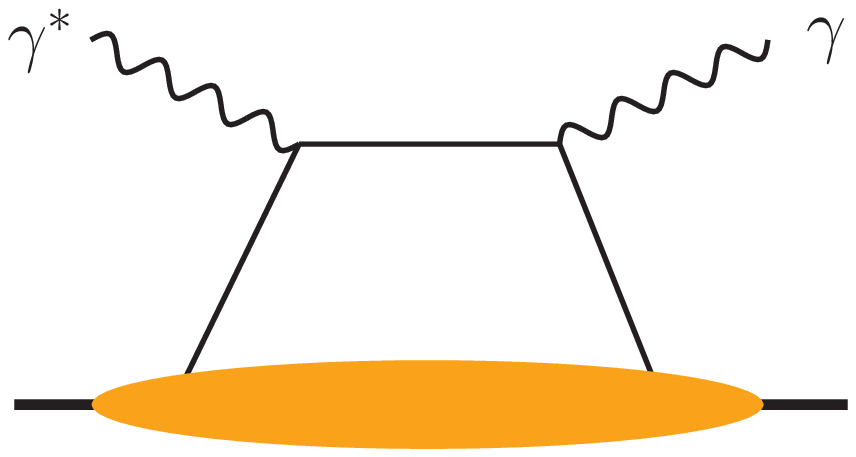}
\hspace{2em}
  \includegraphics[width=0.25\textwidth]{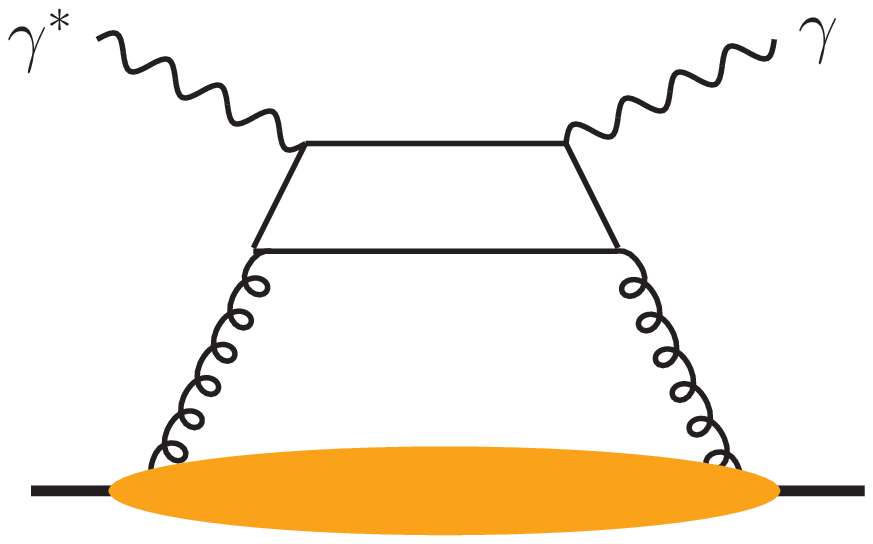}
\hspace{3em}
  \includegraphics[width=0.25\textwidth]{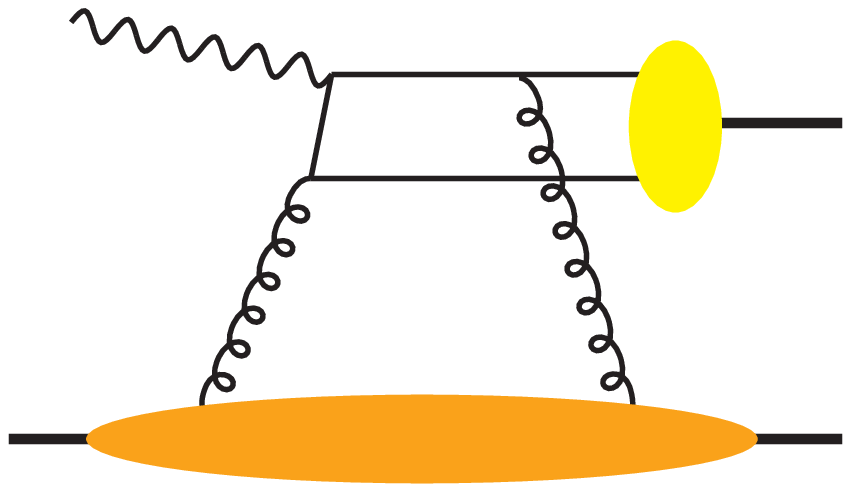}
  \caption{\label{fig:gpd-graphs} Example graphs for DVCS (left and
    centre) and for exclusive vector meson production (right).  The
    lower blob denotes a GPD, and the upper blob in the right panel
    denotes the meson distribution amplitude.}
\end{center}
\end{figure*}

In this section we take a closer look at the theory that connects GPDs
with exclusive scattering processes.  The factorisation of dynamics
into perturbative and non-pertur\-bative quantities happens at the level
of the scattering amplitude.  Schematically, a GPD factorisation
formula reads
\begin{align}
  \label{gpd-fact}
& \mathcal{A}(\xi,\Delta^2,Q^2)
\nonumber \\
 &\quad 
  = \sum_i \int_{-1}^1 dx\, C_i\bigl( x,\xi; \log(Q/\mu) \bigr)
     H_i(x,\xi,\Delta^2; \mu) \,,
\end{align}
where $C_i$ represents the hard-scattering process, $H_i$ is a GPD and
the sum is over the relevant parton types $i$.  Example graphs are
shown in figure~\ref{fig:gpd-graphs}.  The support region of the
integral over $x$ includes the different regimes discussed in the
previous section (see figure~\ref{fig:gpd-bspace}).  In the present
section, we gloss over the fact that there are GPDs with different
dependence on the spin of the partons and the protons.

At lowest order in $\alpha_s$, the hard-scattering kernels for DVCS,
TCS and meson production are linear combinations of
$1/(\xi - x - i\epsilon)$ and $1/(\xi + x - i\epsilon)$.  This gives a
nontrivial convolution for $\re \mathcal{A}$ while $\im \mathcal{A}$
involves $H(x,\xi,\Delta^2)$ at the special points where
$x = \pm \xi$.  At higher orders, logarithms of $Q/\mu$ appear in
$C_i$, where $Q$ is the physical hard scale.  The dependence on the
factorisation scale $\mu$ cancels between $C_i$ and $H_i$ to the order
in $\alpha_s$ considered, just as in the case of inclusive processes
with PDFs.  As already mentioned, the separate dependence of GPDs on
$x$ and $\xi$ cannot be directly inferred from the amplitude: it is
only through the interplay between $x$ and $\mu$ in the scale
evolution of $H_i$ (combined with the dependence of $C_i$ on $Q/ \mu$)
that this aspect of GPDs can be constrained from measurements.  The
evolution equation for GPDs has the form
\begin{align}
& \frac{d}{d\log\mu} \, H_i(x,\xi,\Delta^2;\mu)
\nonumber \\
 &\quad = \sum_j \int_{-1}^1 d\hat{x}\,
    V_{ij}\bigl( x,\hat{x},\xi; \alpha_s(\mu) \bigr)
    H_j(\hat{x},\xi,\Delta^2;\mu) \,.
\end{align}
Notice that evolution changes the $x$ dependence but is local in the
variables $\xi$ and $\Delta^2$, which are fixed by the hadron momenta
in the matrix element defining the GPD.  In the region $|x| > |\xi|$,
evolution acts in a similar way as the familiar DGLAP evolution of
PDFs, whereas for $|x|<|\xi|$ it is similar to the ERBL evolution of
meson distribution amplitudes.  Correspondingly, one often calls the
respective $x$ ranges the DGLAP or ERBL regions.

The evolution kernels $V_{ij}$ are known at two-loop order, matching
the available two-loop accuracy of the hard-scattering coefficients
$C_i$ for light quarks and gluons in DVCS.  One loop corrections to
$C_i$ are available for meson production and for TCS, as well as for
DVCS with a heavy-quark loop.  Detailed information and references can
be found in the review \cite{Belitsky:2005qn}, apart for the more
recent result \cite{Muller:2012yq}.

The hard-scattering mechanism selects certain polarisation
combinations for the photons and/or mesons in the process, so that a
twist-two factorisation formula like \eqref{gpd-fact} only holds for
selected helicity amplitudes.  Other helicity amplitudes are
suppressed by one or two orders in $\Lambda/Q$, where $\Lambda$
represents the scale of non-perturbative dynamics, as well as the
proton mass $m$ and the kinematical scale $|\tvec{\Delta}|$.  In the
case of DVCS and TCS, the leading amplitudes are for transverse
$\gamma^*$ polarisation, whereas those for longitudinally polarised
$\gamma^*$ are suppressed by $|\tvec{\Delta}|/Q$.  They can be
factorised into hard-scattering coefficients and generalised parton
distributions of twist three.  We shall not review these in more
detail here, but mention that certain twist-three distributions are
obtained not from the two-parton correlation function \eqref{corr} but
from its analogue with an additional field operator for a transversely
polarised gluon.

The leading helicity amplitudes have themselves power corrections,
which in the case of Compton scattering are known to be of order
$\Lambda^2/Q^2$, corresponding to twist four.  An important milestone
achieved only recently is the computation of dynamical twist-four
corrections (suppressed by $m^2/Q^2$ or $\tvec{\Delta}^2/Q^2$) for
DVCS \cite{Braun:2014sta}.  At moderate values of $Q^2$ these can be
important in size.

\begin{figure}[hb]
\begin{center}
  \includegraphics[width=0.48\textwidth]{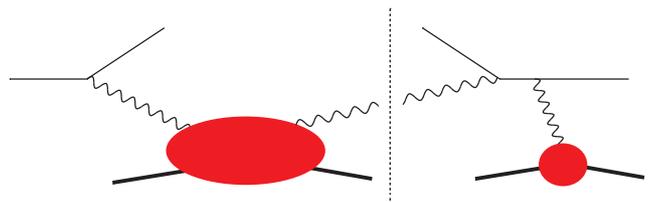}
  \caption{\label{fig:dvcs} Graph for the interference between Compton
    scattering (left) and the Bethe-Heitler process (right) in the
    process $\ell p\to \ell p \gamma$.  The dashed line denotes the
    final state.}
\end{center}
\end{figure}

The helicity of the photon and/or meson leaves an imprint on angular
distributions in the final state.  This provides an experimental
handle to select combinations of helicity amplitudes that appear at
definite order in the $\Lambda/Q$ expansion.  The angular structure of
the cross section is particularly rich in DVCS, because in the
observable process $\ell p\to \ell p \hspace{0.5pt} \gamma$ with
$\ell=e,\mu$, Compton scattering interferes with the Bethe-Heitler
process (see figure~\ref{fig:dvcs}).  The latter can be computed in
QED given the input of the electromagnetic nucleon form factors,
providing a unique tool to probe DVCS and thus GPDs at the amplitude
level \cite{Ji:1996nm,Diehl:1997bu}.  An analogous statement holds for
TCS, $\gamma\ms p \to \ell^+\ell^- p$.

Combining the information from polarisation and angular dependence,
Compton scattering can provide the most detailed access to exclusive
amplitudes \cite{Belitsky:2001ns,Diehl:2005pc}, and the theory to
connect these amplitudes with GPDs is most precise.  On the other
hand, with Compton scattering alone one cannot separate the
distributions for different quark flavors and has only indirect access
to the gluon distributions via $\alpha_s$ effects.

Meson production thus is an important source of complementary
information and offers a variety of experimentally accessible
production channels.  Unfortunately, there is no systematic theory for
power corrections in this case, whose evaluation thus requires some
degree of model dependence.  It remains controversial in the
literature how large such corrections are in experimentally accessible
kinematics, see for instance the discussion in
\cite{Belitsky:2005qn,Diehl:2003ny} and the very different analyses in
\cite{Goloskokov:2007nt} and \cite{Lautenschlager:2013uya}.  A detailed
account of meson production is given later in this
volume~\cite{Favart:2015umi}.

Let us finally mention that there are several detailed reviews on
GPDs.  The early development of the field and its cross connections
with other areas of QCD are discussed in \cite{Ji:1998pc,Goeke:2001tz}
and \cite{Diehl:2003ny}, a wealth of technical material is provided in
\cite{Belitsky:2005qn}, and \cite{Boffi:2007yc} gives a detailed
account of modelling GPDs.


\section{More on TMDs}
\label{sec:TMDs}

Let us now take a closer look at the type of factorisation that
connects observables to TMDs.  Focusing on unpolarised protons for the
moment, the cross section for Drell-Yan production,
$pp\to \ell^+\ell^- + X$ can be written as
\begin{align}
  \label{tmd-fact}
& \frac{d\sigma}{dQ^2\, dY\, d^2\tvec{q}} =
  \frac{1}{s}\, \sum_{ij} \hat{\sigma}_{ij}(Q;\mu)
   \int d^2\tvec{k}_1\, d^2\tvec{k}_2\,
\\
    & \quad \times
\delta^{(2)}(\tvec{q} - \tvec{k}_1 - \tvec{k}_2)\,
        f_i(x_1,\tvec{k}_1;\zeta_1,\mu)\,
        f_j(x_2,\tvec{k}_2;\zeta_2,\mu)
\nonumber
\end{align}
if the transverse momentum $\tvec{q}$ of the lepton pair (and hence of
the virtual photon) is much smaller than is invariant mass $Q$.  The
sum runs over all relevant pairs $(ij)$ of quark and antiquark
distributions.  Notice that in this situation, the longitudinal
momentum fractions of the partons are fixed by external kinematics,
with $x_1 x_2 = Q^2/s$ and $x_1/x_2 = \exp(2Y)$, where $\sqrt{s}$ is
the total collision energy and $Y$ the rapidity of the lepton pair in
the $pp$ centre-of-mass.  This is because the condition
$|\tvec{q}| \ll Q$ suppresses any radiation of unobserved particles
from the hard subprocess.  In the more familiar case where $\tvec{q}$
is integrated over, one has instead
\begin{align}
& \frac{d\sigma}{dQ^2\, dY} =
  \frac{1}{s}\, \sum_{ij} 
  \int_{x_1}^1 \frac{d\hat{x}_1}{\hat{x}_1}
  \int_{x_2}^1 \frac{d\hat{x}_2}{\hat{x}_2}
\\
    & \quad \times
      \tilde{\sigma}_{ij}(\hat{x}_1,\hat{x}_2, Q;\mu)\,
        f_i(x_1/\hat{x}_1; \mu)\, f_j(x_2/\hat{x}_2; \mu) \,,
\nonumber
\end{align}
At leading order in $\alpha_s$, the hard subprocess cross section is
proportional to $\delta(1-\hat{x}_1)\, \delta(1-\hat{x}_2)$, but at
higher orders hard partons are emitted into the final state.  They
carry away longitudinal momentum, so that one probes momentum
fractions in the PDFs from $x_1$ or $x_2$ up to $1$.

The specific form \eqref{tmd-fact} of TMD factorisation has been
established in \cite{Collins:2011zzd}.  It differs in technical
details from the original version of Collins, Soper and Sterman in
\cite{Collins:1981uk,Collins:1984kg} (a brief account of which is
given in \cite{Bacchetta:2008xw}).  Let us highlight the role of the
rapidity regulator parameters $\zeta_1$ and $\zeta_2$ in the TMDs,
which satisfy $\zeta_1 \zeta_2 = Q^4$.  The evolution in these
parameters is simplest if we Fourier transform the TMDs from
$\tvec{k}$ to $\tvec{z}$ (see figure~\ref{fig:scheme}) and then has
the form
\begin{align}
  \label{cs-eq}
\frac{d}{d\log \sqrt{\zeta}}\, f(x,\tvec{z};\zeta,\mu)
 &= K(\tvec{z};\mu)\, f(x,\tvec{z};\zeta,\mu) \,.
\end{align}
The $\mu$ dependence of $f$ is given by a simple renormalisation group
equation
\begin{align}
  \label{mu-eq}
& \frac{d}{d\log\mu}\, f(x,\tvec{z}; \zeta,\mu)
\\
 &\quad = \biggl[ \gamma_f\bigl( \alpha_s(\mu) \bigr)
    - \gamma_K\bigl( \alpha_s(\mu) \bigr)\, \log\frac{\sqrt{\zeta}}{\mu} 
  \ms\biggr]\, f(x,\tvec{z}; \zeta,\mu) \,.
\nonumber 
\end{align}
More detail is given in the contribution \cite{Rogers:2015sqa} to this
volume.  Let us emphasise that TMDs evolve with $\zeta$ and $\mu$ at
\emph{fixed} momentum fraction $x$ -- unlike DGLAP evolution with its
interplay between the $x$ and $\mu$ dependence.  This is directly related
with the absence of longitudinal momentum integrals in the TMD
factorisation formula \eqref{tmd-fact}, which we already explained.  A
convolution integral over longitudinal momentum occurs only when one
computes the high $\tvec{k}$ behaviour of TMDs, as we will see now.

At low $\tvec{k}$ (or equivalently at large transverse distances
$\tvec{z}$ between the quark fields) the TMDs can be thought of as
describing the ``intrinsic'' transverse momentum of partons in the
proton, which arises from non-perturbative dynamics.  Starting from
low transverse momentum and emitting partons with large transverse
momenta, one can obtain values of $\tvec{k}$ much larger than the
scale $\Lambda$ of non-perturbative interactions.  In this situation
one can express TMDs in terms of collinear PDFs and a hard-scattering
kernel describing the emission.  In the transverse-momen\-tum
representation, this reads
\begin{align}
  \label{high-kt-tail}
& f_i(x,\tvec{k}; \zeta,\mu) = \frac{1}{\tvec{k}^2}
\\
 & \quad \times \sum_{j} 
   \int_x^1 \frac{d\hat{x}}{\hat{x}}\, C_{ij}\bigl( \hat{x},
      \log(\zeta /\tvec{k}^2), \log(\mu^2 /\tvec{k}^2)
   \bigr)\, f_j(x/\hat{x}; \mu) \,.
\nonumber 
\end{align}
There is of course a smooth transition between the non-perturbative
regime at low $\tvec{k}$ and the perturbative behaviour in
\eqref{high-kt-tail}, so that there is no strict distinction between
the ``intrinsic'' transverse momentum of a parton and the transverse
momentum generated by hard radiation.

\begin{figure}
\begin{center}
  \includegraphics[width=0.24\textwidth]{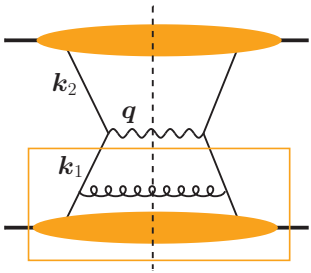}
  \caption{\label{fig:dy-gluon} A lowest-order graph for Drell-Yan
    production at intermediate transverse momentum of the lepton pair.
    The blobs represent collinear PDFs, and the box around the lower
    blob and gluon denotes the TMD \protect\eqref{high-kt-tail} in the
    region of large $\tvec{k}_1$.  The decay of the virtual photon
    into a lepton pair is not shown for simplicity.}
\end{center}
\end{figure}

The form \eqref{high-kt-tail} can be used to compute cross sections in
the regime of intermediate transverse momenta
$\Lambda^2 \ll \tvec{q}^2 \ll Q^2$.  Alternatively, one may describe
the same physics using collinear factorisation, with the hard
subprocess starting at order $\alpha_s$ to allow radiation of at least
one parton with high transverse momentum.  In fact, both descriptions
give the same result at given order in $\alpha_s$, as suggested by
figure~\ref{fig:dy-gluon} and shown for instance in
\cite{Bacchetta:2008xw,Ji:2006ub}.  The formulation using collinear
factorisation has large logarithms of $Q^2/\tvec{q}^2$ in the
hard-scattering cross section.  Their power increases at each order in
$\alpha_s$, which leads to a poor convergence of the perturbative
series.  The formulation starting from TMDs allows one to sum these
so-called Sudakov logarithms to all orders in $\alpha_s$, using the
evolution equations in $\zeta$ and $\mu$ discussed above.  With
reference to the original work \cite{Collins:1984kg}, this is often
called CSS resummation.

The result \eqref{high-kt-tail} clearly shows that the integral
$\int d^2\tvec{k}$ of a TMD requires a suitable regularisation in the
ultraviolet region.  For heuristic purposes, one may think of a simple
cutoff in $\tvec{k}^2$.  Setting this cutoff to $\mu^2$ and taking the
derivative with respect to $\mu^2$, one readily sees that the kernel
$C_{ij}$ in \eqref{high-kt-tail} is closely related to the spitting
kernel in the DGLAP evolution equations for collinear PDFs.  For
systematic calculations, however, one typically defines the collinear
PDFs using dimensional regularisation, subtracting the ultraviolet
divergence in $4-2\epsilon$ dimensions and then setting $\epsilon=0$.
The simple integral relation between TMDs and PDFs defined in $4$
dimensions is then of course lost.  In a modified form, it is however
recovered in the Fourier conjugate representation
\begin{align}
  \label{z-space-tmd}
f_i(x,\tvec{z}; \zeta,\mu) &= \int\!
  d^2\tvec{k}\, e^{i \tvec{k}\tvec{z}}\, f_i(x,\tvec{k}; \zeta,\mu) \,,
\end{align}
where the analogue of \eqref{high-kt-tail} reads
\begin{align}
  \label{small-b-limit}
& f_i(x,\tvec{z}; \zeta,\mu) = f_i(x;\mu)
\\[0.2em]
 &\quad + \sum_{j} \int_x^1 \frac{d\hat{x}}{\hat{x}}\,
    \widetilde{C}_{ij}\bigl(
     \hat{x}, \log(\zeta \tvec{z}^2), \log(\mu^2 \tvec{z}^2) \bigr)\,
   f_j(x/\hat{x}; \mu)
\nonumber 
\end{align}
at small $\tvec{z}$.  The exponential in the Fourier transform indeed
acts as an ultraviolet regulator for the integral, since at high
$\tvec{k}$ its oscillations are sufficient to give a finite result.
The integral of the TMD regulated in this way gives the corresponding
PDF plus corrections that can be computed in an $\alpha_s$ expansion.
The divergence of the unregulated integral is reflected in the
logarithms of $\tvec{z}^2$ on the r.h.s.\ of \eqref{small-b-limit}.
It is amusing to note that for suitable functions $f(x,\tvec{k})$, the
exponential regulator in \eqref{z-space-tmd} is equivalent to a
momentum cutoff \cite{Bacchetta:2008xw}.

\begin{figure}[hb]
\begin{center}
  \includegraphics[width=0.44\textwidth]{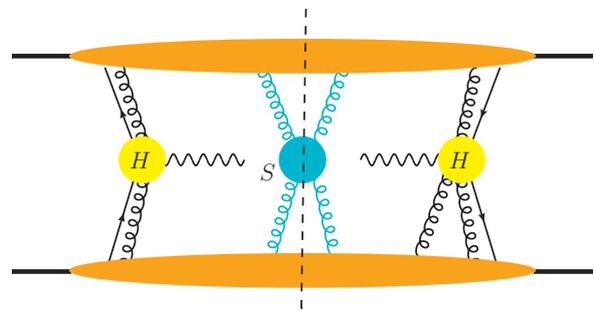}
  \caption{\label{fig:dy-blobs} Organisation of a graph for the
    Drell-Yan process into subgraphs that contain either hard momenta
    ($H$) or soft momenta ($S$) of momenta collinear to one of the
    protons (top and bottom blobs).}
\end{center}
\end{figure}

Let us now take a step back to the derivation of the TMD factorisation
formula \eqref{tmd-fact}.  This formula, and graphs like the one in
figure~\ref{fig:dy-gluon}, suggest that the two protons only
interact via the annihilation of a quark-antiquark pair into a virtual
photon.  This is barely plausible and indeed not the case.  In the
language of perturbation theory, the two protons can exchange an
arbitrary number of soft gluons, and in addition, any number of gluons
with longitudinal polarisation from each can take part in the
$q\bar{q}$ annihilation subprocess, as shown in
figure~\ref{fig:dy-blobs}.  To establish factorisation, one needs to
show that these gluon interactions can be cast into a form consistent
with the simple structure in \eqref{tmd-fact}.  The result of such
arguments, presented in detail in~\cite{Collins:2011zzd} (and sketched
briefly in \cite{Diehl:2015bca}), is that the physical effects of
these gluons are represented by Wilson line operators between the
fields in the parton correlation function \eqref{corr} (integrated
over $k^-$) and by so called soft factors, which are vacuum
expectation values of further Wilson lines and can be absorbed in the
definition of the TMDs.  The Wilson lines also turn the operator
product in~\eqref{corr} into a gauge invariant operator, as is
appropriate for the definition of a meaningful quantity.

All this may seem to be technicalities, but indeed there is important
physics behind it.  The precise form of the Wilson lines allows one to
regulate the rapidity divergences of TMDs, introducing a parameter
$\zeta$.  The associated rapidity evolution equation allows one to
resum large logarithms in physical cross sections, without which one
would badly fail to describe experimentally measured distributions.

\begin{figure*}
\begin{center}
  \includegraphics[width=0.25\textwidth]{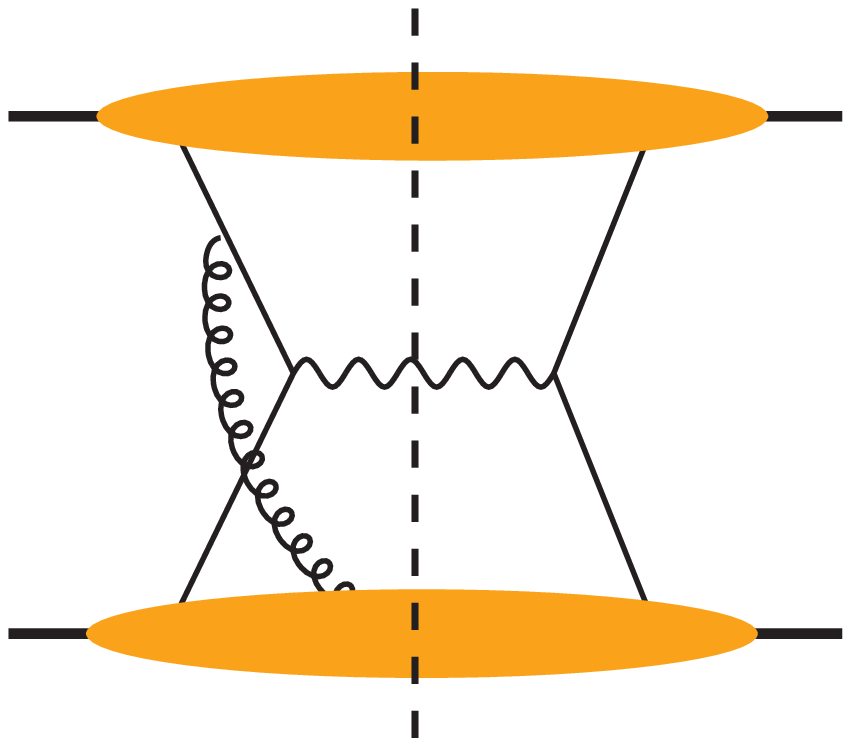}
\hspace{1em}
  \includegraphics[width=0.35\textwidth]{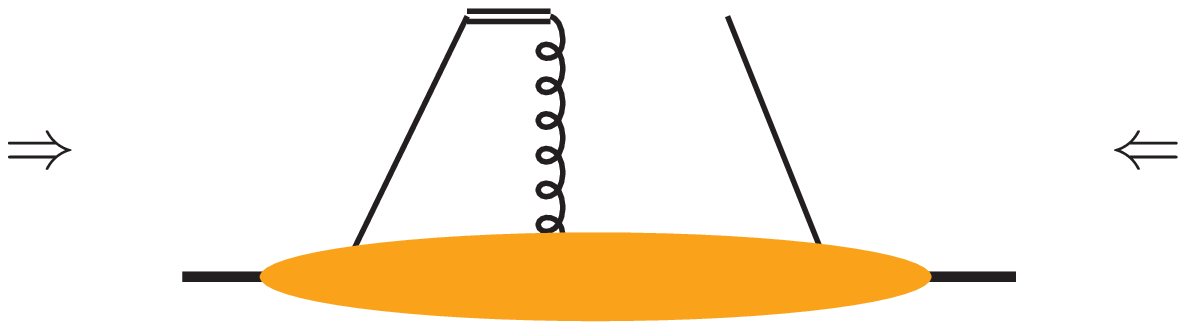}
\hspace{1em}
  \includegraphics[width=0.25\textwidth]{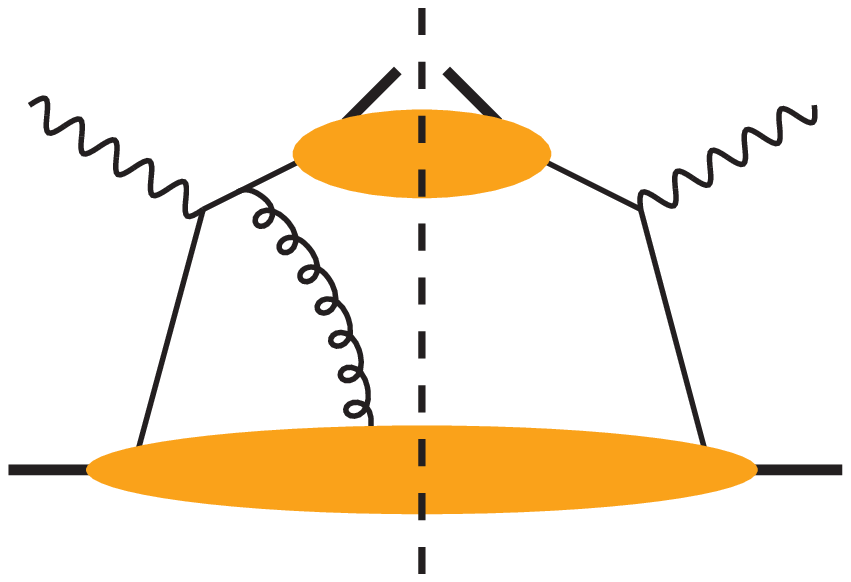}
  \caption{\label{fig:dy-sidis} Gluon exchange graphs in Drell-Yan
    production (left) and semi-inclusive DIS (right), as well as the
    corresponding Wilson line in the definition of the TMD (centre).
    The small blob in the SIDIS graph denotes a transverse-momentum
    dependent fragmentation function.}
\end{center}
\end{figure*}

A far reaching result is that the path of the Wilson lines depends on
the space-time structure of the process in which the TMDs are
embedded.  The Wilson lines required for Drell-Yan production point to
the past, whereas those appearing in the parton distributions for
SIDIS point to the future.  This reflects the fact that the gluon
interactions shown in figure~\ref{fig:dy-sidis} strike a parton
\emph{before} the hard scattering in the Drell-Yan case and
\emph{after} the hard scattering in SIDIS.

This difference has remarkable consequences when spin dependence is
taken into account.  Consider the distribution of unpolarised quarks
in a proton that is polarised in the transverse direction $\tvec{s}$.
For a proton moving in the positive or negative $z$ direction, this
can be parametrised as
\begin{align}
  \label{sivers-def}
& f^{[U]}(x,\tvec{k},\tvec{s})
\nonumber \\
 &\quad = f_1^{[U]}(x,\tvec{k}^2)
  - \frac{\epsilon^{ij} \tvec{k}{}^i \tvec{s}^j}{m}\,
    \text{sign}(P^3)\, f_{1T}^{\perp [U]}(x,\tvec{k}^2) \,,
\end{align}
where $\epsilon^{ij}$ is the antisymmetric tensor in two dimensions.
The superscript $[U]$ indicates the Wilson line dependence; for
simplicity we have omitted the arguments $\zeta$ and $\mu$ in all
functions.  Under time reversal the Wilson lines for Drell-Yan
production turn into the ones for SIDIS, so that the unpolarised
distribution $f_1$ is the same for both processes.  By contrast, the
factor multiplying the Sivers function $f_{1T}^\perp$ in
\eqref{sivers-def} changes sign under time reversal (which flips
both momentum and spin vectors).  Time reversal symmetry thus gives
\begin{align}
  \label{sivers-sign-change}
f_{1T}^{\perp [\text{DY}]}(x,\tvec{k}^2) &=
  - f_{1T}^{\perp [\text{SIDIS}]}(x,\tvec{k}^2) \,.
\end{align}
The modulation of the transverse-momentum distribution induced by
transverse proton polarisation has opposite sign in the two cases -- were
it not for the gluon exchange effects represented by the Wilson line, this
modulation would be zero.  The Sivers distribution (as well as other spin
dependent distributions that are naively zero due to time reversal
invariance) shows in a pointed way that in some situations the
``structure'' of the proton cannot be discussed independently of the
physical process in which this structure manifests itself.

For intermediate transverse momenta,
$\Lambda^2 \ll \tvec{q}^2 \ll Q^2$, one can also compute the graphs in
the left and right panels of figure~\ref{fig:dy-sidis} using collinear
factorisation.  The hadronic input for the proton at the bottom of the
graphs is then a twist-three distribution $T_F(x_1,x_2)$, called
Qiu-Sterman function.  The large $\tvec{k}$ limit of the Sivers
distribution can be expressed in terms of this function, as well.  In
Fourier space, the result can be cast into the form\footnote{The
  relation \protect\eqref{small-b-sivers} can be obtained from
  eq.~(47) in \protect\cite{Aybat:2011ge}.}
\begin{align}
  \label{small-b-sivers}
& \int\! d^2\tvec{k}\, e^{i \tvec{k}\tvec{z}}\,
   \frac{\tvec{k}^2}{m}\,
   f_{1T}^{\perp [\text{SIDIS}]}(x,\tvec{k}; \zeta,\mu)
\nonumber \\
 &\hspace{8em} = - T_F(x,x;\mu) + \mathcal{O}(\alpha_s) \,,
\end{align}
which relates a regulated $\tvec{k}$ integral of the Sivers
distribution $f_{1T}^{\perp}(x,\tvec{k}^2) $ with the Qiu-Sterman
function, in analogy to the relation \eqref{small-b-limit} for
unpolarised distributions.  Using \eqref{sivers-sign-change} and
\eqref{small-b-sivers} to calculate the Sivers asymmetries in
Drell-Yan production and in SIDIS, one obtains agreement with the
collinear twist-three calculation at intermediate transverse momenta
\cite{Ji:2006ub,Ji:2006br}.  A general analysis of the relation
between the two formalisms for a large class of spin asymmetries is
given in \cite{Bacchetta:2008xw}.

It is not always possible to describe the effects of soft gluon
exchange by Wilson line operators and to obtain a factorisation
formula with TMDs.  Processes for which TMD factorisation has been
established are SIDIS and $e^+e^-$ annihilation into back-to-back
hadrons (both involving trans\-verse-momentum dependent fragmentation
functions), as well as hadron-hadron collisions in which only
colourless particles are produced by the hard scattering (e.g.\ a
virtual photon, a $\gamma\gamma$ pair, a $Z$ or $W$, a Higgs boson,
etc.).  For hadron-hadron collisions with observed hadrons in the
final state, soft gluon exchange between the two hadrons generically
breaks TMD factorisation \cite{Rogers:2010dm,Rogers:2013zha}.  Because
soft gluon interactions cannot be reliably computed in perturbation
theory, it is difficult to predict how large such factorisation
breaking effects are.


\section{Spin and orbital angular momentum}
\label{sec:spin}

Both TMDs and GPDs have a rich structure in the parton and proton spin.
They can in particular express correlations between transverse momentum or
position and transverse polarisation.  An example is the Sivers function
we already encountered in the previous section.  It is instructive to
compare the transverse-momentum distribution \eqref{sivers-def} with the
impact parameter distribution of unpolarised quarks in a transversely
polarised proton, given by~\cite{Burkardt:2002hr}
\begin{align}
  \label{impact-spin-corr}
& f(x,\tvec{b},\tvec{s})
\nonumber \\
 &\quad = H(x,\tvec{b}^2) + \frac{\epsilon^{ij}\tvec{b}{}^i \tvec{s}^j}{m}\, 
  \text{sign}(P^3)\,
  \frac{\partial}{\partial\tvec{b}^2} E(x,\tvec{b}^2) \,,
\end{align}
where $H(x,\tvec{b}^2)$ and $E(x,\tvec{b}^2)$ are obtained by a
two-dimen\-sional Fourier transform from the GPDs $H(x,\xi,\Delta^2)$ and
$E(x,\xi,\Delta^2)$ at $\xi=0$.  In contrast to the case of the Sivers
function, the factor multiplying $E$ in \eqref{impact-spin-corr} is not
time reversal odd, because time reversal flips spins and momenta but not
spatial directions.  The integral of $E(x,\tvec{b}^2)$ over $x$ is related
to the electromagnetic Pauli form factor of the nucleon and thus rather
well known experimentally.

The decomposition \eqref{impact-spin-corr} shows that transverse proton
polarisation induces a sideways shift in the impact parameter distribution
of the struck quark, which implies a shift of the spectator partons in the
opposite direction (so that the overall centre of momentum remains
unchanged).  A fruitful idea is to relate this spatial anisotropy to an
anisotropy in the transverse-momentum distribution of the struck quark: as
we have already discussed, the Sivers asymmetry owes its very existence to
the interactions involving the spectator partons in the proton (see
figure~\ref{fig:dy-sidis}).  This mechanism has been called
\emph{chromodynamic lensing} \cite{Burkardt:2003uw} and connects in
particular the sign of the Sivers distribution with the sign of the
anomalous magnetic moment, in agreement with phenomenology.  To quantify
this relation is difficult, given the non-perturbative nature of spectator
parton interactions.  In specific models, using for instance perturbative
gluon exchange and representing the spectator system by a diquark, one
obtains however definite relations between TMDs and GPDs, as discussed for
instance in \cite{Meissner:2007rx}.

Another feature of the distribution $E(x,\xi,\Delta^2)$ becomes evident if
one changes basis from transversely polarised proton states to
longitudinally polarised ones.  One then finds that (in contrast to its
unpolarised counterpart $H$), the distribution $E$ contributes to
transitions that reverse the helicity of the proton while preserving the
helicity of the quarks (see e.g.\ \cite{Diehl:2003ny}).  Since the total
angular momentum $J_z$ along $z$ is conserved, orbital angular momentum
must be transferred, which is possible because $\tvec{\Delta}$ is nonzero.
Specifically, one can show that $E$ involves the interference between
light-cone wave functions $\psi$ and $\psi^*$ that differ by one unit of
orbital angular momentum $L_z$ along $z$ \cite{Burkardt:2005km}.

A seemingly different connection between $E$ and angular momentum is given
by Ji's sum rule \cite{Ji:1998pc,Ji:1996ek}
\begin{align}
  \label{ji-rule}
J^q(\mu) = \frac{1}{2} \int_{-1}^1 dx\, x
 & \bigl[ H^q(x,\xi,\Delta^2;\mu)
\nonumber \\
 & + E^q(x,\xi,\Delta^2;\mu) \bigr]_{\Delta^2 = 0} \,.
\end{align}
Here $J^q(\mu)$ is the \emph{total} angular momentum along $z$ carried by
quarks and antiquarks of flavour $q$ in a proton polarised along $z$.  It
is defined via the matrix element of an appropriate angular momentum
operator between proton states and includes both helicity and orbital
contributions.  We note that the prefactor of $E$ in
\eqref{impact-spin-corr} can be rewritten in terms of the cross product
between $\tvec{b}$ and the proton momentum, which is indicative of orbital
angular momentum.  The relation between this and the sum rule
\eqref{ji-rule} is rather subtle and has been discussed in
\cite{Burkardt:2005hp}.

To determine from experimental information on GPDs is extremely
demanding, not only because the sum rule requires extrapolation to
$\Delta^2 = 0$ but also due to the problem of reconstructing the $x$
dependence of GPDs (see section \ref{sec:GPDs}).  By contrast, it is
comparatively straightforward to evaluate $J^q(\mu)$ in lattice QCD,
and there is indeed considerable activity in this direction
\cite{Hagler:2009ni,Liu:2015xha}.

It is important to note that there are several distinct ways to decompose
the total angular momentum of the proton into contributions $J^q$ and
$J^g$ from quarks and gluons, and to further separate these into
contributions from orbital angular momentum and from intrinsic spin.
Apart from the decomposition in \cite{Ji:1996ek}, which is used in the sum
rule \eqref{ji-rule}, there is in particular the operator definition by
Jaffe and Manohar \cite{Jaffe:1989jz}.  Several other operator definitions
have been proposed, and the contribution \cite{Liu:2015xha} to this volume
gives a detailed discussion of this issue.

Various decompositions of angular momentum differ by terms involving the
gluon field.  This points to the subtle nature of gauge interactions -- a
theme we encountered already when discussing the role played by Wilson
lines in the definition of TMDs.  In a broader context, one should also
remember that there may not always be a unique way to promote quantities
from classical mechanics to quantum mechanics, let alone to quantum field
theory.  Indeed, even the definition of the quark helicity contribution
the proton spin is subject to ambiguities due to the axial anomaly (see
for instance \cite{Lampe:1998eu}).  The existence of such quantum effects
may complicate a physical interpretation, but one should keep in mind that
when calculating physical observables, a correct answer can be obtained
using different schemes.


\begin{acknowledgement}
It is a pleasure to thank Andreas Sch\"afer and Frank Tackmann for
valuable input.
\end{acknowledgement}


\end{document}